\documentclass[letterpaper, 10 pt, conference]{ieeeconf}  

\IEEEoverridecommandlockouts                          

\usepackage{eso-pic}
\usepackage{graphics} 
\usepackage{epsfig} 
\usepackage{amsmath} 
\usepackage{amssymb}  
\usepackage{graphicx}
\usepackage{bm}
\usepackage{multirow}
\usepackage{makecell}
\usepackage{BernsteinStyle}
\usepackage{cite}
\usepackage{tikz}
\usepackage{longtable}
\usepackage{tabularx}
\newcolumntype{C}[1]{>{\centering\let\newline\\\arraybackslash\hspace{0pt}}m{#1}}

\usepackage{hhline}
\usetikzlibrary{calc}
\usetikzlibrary{shapes, arrows}
\usetikzlibrary{positioning}
\usepackage{circuitikz}
\usetikzlibrary{patterns,decorations.pathmorphing,positioning}
\usepackage[linesnumbered,ruled,vlined,noend]{algorithm2e}
\newcounter{example}
\newenvironment{example}[1][]{\refstepcounter{example}\par\medskip
   \noindent \textbf{\indent Example~\theexample. #1} \rmfamily}{\medskip}

\usetikzlibrary{shapes,arrows,calc,positioning}
\tikzstyle{bigblock} = [draw, fill=blue!20, rectangle, 
    minimum height=6em, minimum width=8em]
\tikzstyle{medblock} = [draw, fill=blue!20, rectangle, 
    minimum height=4em, minimum width=4em]    
\tikzstyle{mux} = [draw, fill=black!20, rectangle, 
    minimum height=5em, minimum width=0.1em]    
\tikzstyle{smallblock} = [draw, fill=blue!20, rectangle, 
    minimum height=2.5em, minimum width=4em]
\tikzstyle{sum} = [draw, fill=blue!20, circle, node distance=1cm]
\tikzstyle{signal} = [coordinate]
\tikzstyle{pinstyle} = [pin edge={to-,thin,black}]
\tikzstyle{block} = [draw, fill=blue!20, rectangle, 
    minimum height=3em, minimum width=6em]
\tikzstyle{blockS} = [draw, fill=blue!20, rectangle, 
    minimum height=3em, minimum width=4em]  
\tikzstyle{sum} = [draw, fill=blue!20, circle, node distance=1.5cm]
\tikzstyle{gain} = [draw, fill=blue!20, regular polygon, regular polygon sides = 3, node distance=1.25cm, shape border rotate = -90]
\tikzstyle{mult} = [draw, fill=blue!20, circle, inner sep=0pt, minimum size=0.2cm,]
\tikzstyle{input} = [coordinate]
\tikzstyle{output} = [coordinate]
\tikzstyle{ground}=[fill,pattern=north east lines,draw=none,minimum width=0.75cm,minimum height=0.3cm]

\title{\LARGE \bf
Nonlinear Predictive Cost Adaptive Control of Pseudo-Linear 
Input-Output Models Using Polynomial, Fourier, and Cubic Spline Observables
}

\author{  Rami Abdulelah Alhazmi$^{1}$,
Achinth Suresh Babu$^{2}$, Syed Aseem Ul Islam$^{2}$, and Dennis S. Bernstein$^{2}$
    \thanks{*This work was supported by King Fahd University of Petroleum and Minerals and NSF under grant CMMI 2310300.}
    \thanks{$^{1}$Rami Abdulelah Alhazmi is with the Department of Aerospace Engineering, King Fahd University of Petroleum and Minerals, Dhahran, Saudi Arabia, and with the Department of Aerospace Engineering, University of Michigan, Ann Arbor, MI, USA
    {\tt\small rahazmi@umich.edu}}%
    \thanks{$^{2}$Achinth Suresh Babu, Syed Aseem Ul Islam, and Dennis S. Bernstein are with the Department of Aerospace Engineering, University of Michigan, Ann Arbor, MI, USA
    {\tt\small \{achinth, aseemisl, dsbaero\}@umich.edu}}%
}

\newcommand{\addcopyrightnotice}{%
  \AddToShipoutPictureFG*{%
    \AtPageLowerLeft{%
      \raisebox{0.5in}{%
        \hspace{0.55in}%
        \parbox{\dimexpr\paperwidth-1.10in\relax}{%
          \centering\scriptsize
          \textcopyright~2026 IEEE. Personal use of this material is permitted.
          Permission from IEEE must be obtained for all other uses, in any
          current or future media, including reprinting/republishing this
          material for advertising or promotional purposes, creating new
          collective works, for resale or redistribution to servers or lists,
          or reuse of any copyrighted component of this work in other works.
        }%
      }%
    }%
  }%
}

\begin{document}
\setlength{\abovedisplayskip}{2pt}
\setlength{\belowdisplayskip}{2pt}
\setlength{\abovecaptionskip}{-4pt}
\setlength{\textfloatsep}{8pt}

\maketitle
\addcopyrightnotice
\thispagestyle{empty}
\pagestyle{empty}

\begin{abstract} 
Control of nonlinear (NL) systems with high levels of uncertainty is practically relevant and theoretically challenging.
This paper presents a numerical investigation of an adaptive NL model predictive control (MPC) technique that relies entirely on online system identification without prior modeling, training, or data collection.
In particular, the paper extends predictive cost adaptive control (PCAC) for linear systems, which is an extension of generalized predictive control, to NL systems.
NL PCAC (NPCAC) uses recursive least squares (RLS) with subspace of information forgetting (SIFt) to identify a discrete-time, pseudo-linear, input-output model, which is used with iterative MPC for NL receding-horizon optimization.
The performance of NPCAC is illustrated using polynomial, Fourier, and cubic-spline basis functions.
\end{abstract}


\section{Introduction}
Model predictive control (MPC) is second only to PID as a successful control technique \cite{samadetal}.
MPC determines control inputs via online, receding-horizon optimization using a model, applying only the first component to the system before repeating the process.
Given an accurate model, MPC succeeds by stabilizing systems while enforcing control constraints, such as magnitude and rate saturation, and state constraints, such as safety limits on position and temperature \cite{kwonpearson,keerthi1988optimal,kwon2006receding,camacho2013model,rawlings2017model}.
For linear models, receding-horizon optimization can be solved using the backward Riccati recursion \cite{kwon2006receding} or quadratic programming (QP), where the latter enforces control and state constraints
For systems with NL dynamics, various techniques have been developed for NL MPC.
For example, symbolic numerical differentiation is used in CasADi \cite{andersson2019casadi} together with iterative methods, such as interior-point methods \cite{wright2005interior}, sequential QP \cite{Boggs_Tolle_1995,PhilipSQP2012}, and quasi-linear parameter varying \cite{Herbert,HerbertConvergence}.
Although early applications of MPC focused on process control \cite{morarireview}, the advent of fast embedded computing has extended the reach of MPC to aerospace systems \cite{eren2017model}.

MPC also lends itself to adaptive control, where system identification is performed concurrently online as a form of indirect adaptive control.
This approach is embodied by generalized predictive control (GPC) \cite{ClarkeGPC1,ClarkeGPC2,Bitmead:90,Mosca:95}, which uses MPC with online system identification performed by recursive least squares (RLS) with variable-rate forgetting (VRF).
Predictive cost adaptive control (PCAC) \cite{islamPCAC} extends GPC by using QP to enforce output constraints and windowed VRF to accelerate learning within RLS.
PCAC was applied experimentally in \cite{rezaMSSP,rileyQCACC2025}
as well as numerically in  
\cite{rileyBACT2025JGCD,aseemF16JGCD}.
In many of these applications, PCAC is effective for NL systems despite the fact that linear models are identified.
Henceforth, in the present paper, we refer to PCAC as linear PCAC (LPCAC), to distinguish it from the NL extension discussed below.
For online NL identification and MPC, the Koopman approach is used in \cite{Williams2015,HerbertKoopRLS,KORDA2018149}.

The goal of the present paper is to extend the effectiveness of LPCAC to more challenging NL systems; this is done by identifying a NL model and solving a nonconvex optimization problem.
For NL PCAC (NPCAC), we consider discrete-time (DT), input-output (IO) models whose coefficients may depend on the output; these NL models are thus {\it pseudo-linear} (PL) as considered in \cite{Balbis,Kamaldar_2023,MPG,alhazmiIMPCACC2025}.
For online system identification, NPCAC uses RLS with subspace of information forgetting (SIFt) to identify a DT, PL, IO model \cite{lai2024siftrls}.
For receding horizon optimization, we use iterative model predictive control (IMPC) as described in \cite{alhazmiIMPCACC2025}.
IMPC is chosen due to the fact that it requires no linearization and is numerically reliable, at least for the examples considered in this paper.
Since IO models involve no internal state, NPCAC requires only the model outputs;  hence, NPCAC is an adaptive output feedback controller. 
Finally, although NPCAC is based on DT models, it can be applied to continuous-time systems under sampled data control, as in the case of LPCAC  \cite{islamPCAC,rezaMSSP,rileyQCACC2025,rileyBACT2025JGCD,aseemF16JGCD}.


\section{ Pseudo-Linear Input-Output System} 
The present paper considers the discrete-time pseudo-linear (PL) scalar input-output (IO) system 
\begin{gather}
    {y}_{k} = \sum_{i=1}^{n}-{F}_{i}(Y_{k-i:k-n}) y_{k-i} 
    +{G}_{i}(Y_{k-i:k-n}) u_{k-i} 
    , \label{eq:ydef}
\end{gather}
where, 
for all $k\ge0$, $y_k\in\BBR$ and $u_k\in\BBR$,
$n\ge1$ is the model order, and,
for all $i=1,\ldots,n$,
\begin{gather}
    Y_{k-i:k-n}\isdef \matl y_{k-i} & \cdots & y_{k-n} \matr \in 
    \BBR^{1\times (n+1-i)},
    \\
    {F}_{i} \colon \BBR^{1\times (n+1-i)} \to \BBR,\quad 
    {G}_{i} \colon \BBR^{1\times (n+1-i)} \to \BBR.
\end{gather}

\section{Online Identification using RLS} 
In this section, the PL IO (PLIO) model to approximate the unknown system \eqref{eq:ydef} is defined.
Then, the recursive least squares (RLS) method used for online identification of the PLIO model is presented.

Let $\hat n\ge 1$, $\ell_f\ge1$, $\ell_g\ge1$,
and, for all $k\ge 0,$ let 
$\bar{F}_{1,k},\ldots,\bar{F}_{\hat n,k}\in\BBR^{1\times \ell_{f}}$
and 
$\bar{G}_{1,k},\ldots,\bar{G}_{\hat n,k}\in\BBR^{1\times \ell_{g}}$
be the coefficient row vectors to be estimated using RLS.
Furthermore, for all $k\ge 0,$ let $\hat y_k\in\BBR$ be an estimate of $y_k$ defined by
\begin{align}
    \hat{y}_{k} &\isdef \sum_{i=1}^{\hat n}-{\bar F}_{i,k} f_{i,k} y_{k-i} 
    + \bar{G}_{i,k}g_{i,k} u_{k-i} 
    , \label{eq:yhat}
\end{align}
where, for $i = 1,\ldots,\hat n$, 
\begin{gather}
    f_{i,k} \isdef f_{i}(Y_{k-i:k-\hat n})\in \BBR^{\ell_{f}},\ 
    f_{i}\colon \BBR^{\hat n+1-i } \to \BBR^{\ell_{f}},\\ 
    g_{i,k} \isdef g_{i}(Y_{k-i:k-\hat n})\in \BBR^{\ell_{g}},\
    g_{i}\colon \BBR^{\hat n+1-i } \to \BBR^{\ell_{g}},
\end{gather}
are vectors of basis functions.
Note, 
for all $k\ge 0$ and $i=1\ldots,\hat n$, let $\widehat{F}_{i,k}\colon\BBR^{1\times\hat n +1-i}\to\BBR$ and $\widehat{G}_{i,k}\colon\BBR^{1\times\hat n +1-i}\to\BBR$ be approximate functions of $F_i$ and $G_i$ defined by
\begin{gather}
    \widehat{F}_{i,k}(Y_i) \isdef {\bar F}_{i,k} f_{i}(Y_i),\quad 
    \widehat{G}_{i,k}(Y_i) \isdef {\bar G}_{i,k} g_{i}(Y_i),
\end{gather}
where $Y_i\in\BBR^{1\times\hat n +1-i}$.
For all $k\ge 0$, the PLIO model \eqref{eq:yhat} can be written as
\begin{align}
    \hat y_k = \bar\theta_k \phi_k , \label{eq:thetaphi}
\end{align}
where
\begin{align} 
    \bar\theta_k &\isdef 
    \matl 
    \bar F_{1,k} & \cdots & \bar F_{\hat n,k} & 
    \bar G_{1,k} & \cdots & \bar G_{\hat n,k} 
    \matr
    \in \BBR^{1\times \ell_\phi}, \label{eq:thetadef}  \\
    \phi_k &\isdef 
    \renewcommand{\arraystretch}{0.8}
    \matl 
    -f_{1,k} y_{k-1} \\ \vdots \\ -f_{\hat n,k} y_{k-\hat n} \\
    g_{1,k} u_{k-1} \\ \vdots \\ g_{\hat n,k} u_{k-\hat n} 
    \matr \in \BBR^{ \ell_\phi},\quad
    \ell_\phi \isdef  \hat n ( \ell_{f} + \ell_{g} ).
\end{align}
For online identification, RLS is used to estimate the coefficients of the IO model \eqref{eq:yhat}.
Specifically, we use the subspace of information forgetting RLS (SIFt-RLS) proposed by \cite{lai2024siftrls}.
The first step in SIFt-RLS is the information filtering, which consists of 
\begin{gather} 
    \bar y_k =
    \left\{
    \setlength{\arraycolsep}{2pt} 
    \hbox{\footnotesize
    $\begin{array}{ll}
        y_k, & \Vert \phi_k\Vert>\varepsilon,\\
        {0}, & {\rm{else}},
    \end{array}$
    }
    \right. \quad
    \bar\phi_k =
    \left\{
    \setlength{\arraycolsep}{2pt} 
    \hbox{\footnotesize
    $\begin{array}{ll}
        \phi_k, & \Vert\phi_k\Vert>\varepsilon,\\
        {0}_{\ell_\phi\times 1}, &{\rm{else}},
    \end{array}$
    }
    \right.
\end{gather}
where $\varepsilon>0$ is a tuning parameter ensuring numerical stability. 
Then, the forgetting is applied to the information subspace. In other words, the forgetting is applied to the covariance subspace parallel to the direction of the regressor.
Let $R_0 \in\BBR^{\ell_\phi \times \ell_\phi}$ be a positive definite covariance matrix, and $P_0 \isdef R_0^{-1}$. Then,
\begin{align}
    \bar R_k &=  R_k - (1-\lambda) R_k \bar\phi_k (\bar\phi_k^\rmT R_k \bar\phi_k )^{-1} \bar\phi_k^\rmT R_k ,\\
    \bar P_{k} &= P_{k} + \frac{1-\lambda}{\lambda} \bar\phi_k (\bar\phi_k^\rmT R_k \bar\phi_k)^{-1}\bar\phi_k^\rmT,
\end{align}
where $\lambda\in(0,1]$ is the forgetting factor.
Lastly, the update step consists of
\begin{align}
    R_{k+1} &= \bar R_k + \bar\phi_k^\rmT \bar\phi_k,\\
    P_{k+1} &= \bar P_k - \bar P_k \bar\phi_k (1+\bar\phi_k^\rmT \bar P_k \bar\phi_k)^{-1}\bar\phi_k^\rmT \bar P_k ,\\
    \bar\theta_{k+1} &= \bar\theta_{k}+    ( \bar y_{k}  -  \bar\theta_{k}  \bar\phi_k  )\bar\phi_k^\rmT P_{k+1}. \label{eq:RLS3}
\end{align} 
$\bar\theta_{0}\in\BBR^{1 \times \ell_\phi}$ is the initial coefficient matrix.
Note that $\bar\theta_{k+1}$, computed using  \eqref{eq:RLS3}, is available at step $k,$ and thus, 
$\bar{F}_{1,k+1},\ldots,\bar{F}_{\hat n,k+1}, \bar{G}_{1,k+1},\ldots,\bar{G}_{\hat n,k+1}$ are available at step $k$ by the definition of $\bar\theta_k$ in \eqref{eq:thetadef}.
%
%

%

\section{Model Predictive Control}
Let $\ell\ge1$ be the preceding horizon number of steps, and 
for all $k\ge 0 $ and $i = 1,\ldots,\ell$, the predicted counterpart sequences of all signals are denoted by $\chi_{k|i} \isdef \chi_{k+i}$,
where 
$\chi_{k|i}$ is the value of a signal $\chi$ at step $(k+i)$, using the information available at step $k$.
At step $k\ge\hat n$, the model predictive control (MPC) optimized cost is given by
\begin{align}
    J_k &\isdef 
    \sum_{i=2}^{\ell+1} Q (r_{k|i}-y_{k|i})^2 + \sum_{i=1}^{\ell}   R u_{k|i}^2
    , \label{eq:qp_cost}
\end{align}
where
$Q\in \BBR$ is the positive-semi-definite command-following weights, 
$R\in \BBR$ is the positive definite control weights, and,
for all $i=1,\ldots,\ell,$
$r_{k|i}$ is the command signal.
The optimized cost \eqref{eq:qp_cost} was subject to the identified model \eqref{eq:yhat}.
Then, $u_{k+1} = u_{k|1}$.
For $k = 1, \ldots, \hat n $, $u_{k+1}$ is sampled from $\mathcal{N}(0,\sigma_u )$, where $\sigma_u>0$ is a tuning parameter.
We assume for all $k\ge \hat n$, $r_k$ is known to the controller.
We define the command-following error by
$e_{{\rm c},k} \isdef r_k - y_k$.

\subsection{MPC For Linear Models}\label{LPCAC}
Let, for $i=1,\ldots,\hat n$, $f_{i} = 1$ and $g_{i} = 1$.
%
Then, the identified model \eqref{eq:yhat} is linear and a linear MPC solver can be used to compute $u_{k+1}$, such as in linear predictive cost adaptive control (PCAC). 

\subsection{MPC For Nonlinear Models}

This section presents iterative MPC (IMPC) for controlling PLIO models.
At step $k$, IMPC computes a sequence of control inputs and predicted outputs over the horizon of length $\ell \ge 1$.
To do this, a subiteration is performed $j=0,\ldots,\nu$ times, where $\nu\ge0$ is the maximum number of subiteration.
At subiteration $j$, the state-dependent-coefficients (SDC's) are constructed using the predicted sequence of outputs and controls, 
which are used to construct quadratic programming (QP),
which minimizes the quadratic cost to obtain a predicted sequence of controls,
which are used to construct a predicted sequence of outputs.
Lastly, the Broyden root-finding technique computes a new sequence of controls to accelerate the convergence of the subiteration.
Finally, at $j=\nu$, the first component of the optimized sequence of controls is applied at step $k+1$.

\subsubsection{Prediction over the Horizon}
At step $k$, IMPC takes the measurements $y_k$ and applied control $u_k$, and computes the next control $u_{k+1}$. 
Hence, IMPC timing takes into account the computational delay \cite{alhazmiIMPCACC2025}.

To compute $u_{k+1}$, first, the one-step ahead output is predicted
\begin{align}
    \hat y_{k+1} &= \bar\theta_{k+1} \phi_{k+1}.\label{eq:y1kdef}
\end{align}
Next, for all $k\ge0,$ $i = 1,\ldots,\ell,$ and $j = 0,\ldots,\nu,$ $\chi_{k|i,j}$ denotes the predicted value of $\chi_{k+i}$ at step $k$ and subiteration $j$.
Thus, for all $i = 2,\ldots,\ell+1$ and $j = 0,\ldots,\nu,$ the predicted outputs are given by
\begin{align}
    y_{k|i,j} &= \hat\theta_{k|i,j}\varphi_{k|i,j}, \label{DNL_SCDC}
\end{align}
where
\begin{align} 
    \hat\theta_{k|i,j} &\isdef 
    \matl 
    \Tilde F_{1,k|i,j} & \cdots & \Tilde F_{\hat n,k|i,j} &
    \Tilde G_{1,k|i,j} & \cdots & \Tilde G_{\hat n,k|i,j}
    \matr
    \nn\\&\quad 
    \in \BBR^{1\times \ell_\varphi}, 
\end{align}%
\begin{align} 
    \varphi_{k|i,j}&\isdef  
    \setlength{\arraycolsep}{4pt}
    \matl 
    y_{k|i-1,j} & \cdots &  y_{k|i-\hat n,j} & 
    u_{k|i-1,j} & \cdots &  u_{k|i-\hat n,j} 
    \matr^\rmT 
    \nn\\&\quad 
    \in \BBR^{ \ell_\varphi},
\end{align}
where $\ell_\varphi \isdef 2\hat n$, and, for all $\iota = 1,\ldots,\hat n$,
\begin{align}
    \Tilde F_{\iota,k|i,j} &\isdef -\widehat F_{\iota,k}(Y_{k+i-\iota:k+i-\hat n,j}), \label{ABCelldef_1}\\
    \Tilde G_{\iota,k|i,j} &\isdef \widehat G_{\iota,k}(Y_{k+i-\iota:k+i-\hat n,j}), \label{ABCelldef_2}\\
    Y_{k+i-\iota:k+i-\hat n,j} &\isdef \matl y_{k|i-\iota,j} & \cdots & y_{k|i-\hat n,j} \matr,
\end{align}
for all $j = 0,\ldots,\nu,$ $y_{k|1,j} \isdef \hat y_{k+1}$, 
for all $\iota=0,\ldots,\hat n-1$,
$y_{k|-\iota,j} \isdef y_{k-\iota}$.
For $j = 0,\ldots,\nu,$ we write \eqref{DNL_SCDC}, for all $i=2,\ldots,\ell+1$, 
\begin{align}
    y_{k|2,j} &= \hat\theta_{k|2,j}\varphi_{k|2,j}, \label{equ:yk_1}\\
    %
    &\vdots \nn\\
    y_{k|\ell+1,j} &=  \hat\theta_{k|\ell+1,j}\varphi_{k|\ell+1,j}. \label{equ:ykell}
\end{align}
Furthermore, we define the output, control, and command windows
\begin{gather}  
    Y_{k,j} \isdef
    \renewcommand{\arraystretch}{0.5}
    \matl
    y_{k|2,j} \\ \vdots \\ y_{k|\ell+1,j} \matr \in \BBR^{ \ell}, \
    U_{k,j} \isdef
    \renewcommand{\arraystretch}{0.5}
    \matl
    u_{k|1,j} \\ \vdots \\ u_{k|\ell,j} \matr \in \BBR^{ \ell}, \label{def_X_U}\\
    \mathcal{R}_{k} \isdef
    \renewcommand{\arraystretch}{0.5}
    \matl 
    r_{k|2} \\ \vdots \\ r_{k|\ell+1} \matr \in \BBR^{ \ell}, \
    \SY_{k,j} \isdef \matl Y_{k,j} \\ U_{k,j} \matr \in \BBR^{ 2 \ell }. \label{XUdef}    
\end{gather}

\subsubsection{Cost and Constraints} 
We define the IMPC cost at step $k$
\begin{gather}
    J_k(\SY_{}) \isdef \SY_{}^\rmT\, \SH \, \SY_{} + \SF_k^\rmT \SY  , \label{f_cost}
\end{gather}
where
\begin{gather}
    \SH \isdef {\rm diag}(I_\ell \otimes Q,I_\ell \otimes R) \in \BBR^{2\ell\times 2\ell},\label{hdef}\\
    \SF_k \isdef \matl -2(I_\ell \otimes Q)\mathcal{R}_{k}\\0_{\ell\times 1} \matr \in \BBR^{2\ell},
\end{gather}
$\SY  \in \BBR^{ 2 \ell }$ is the optimization variable, and $\otimes$ denote the Kronecker product.
The cost in \eqref{f_cost} is subject to equality constraints that arise from \eqref{DNL_SCDC}.
To formulate these constraints, we rewrite \eqref{equ:yk_1}--\eqref{equ:ykell} as
\begin{align}
    Y_{k,j} &= 
    \setlength{\arraycolsep}{2pt}
    \matl {F}_{d,k,j} &{G}_{d,k,j}\matr D_k+ {F}_{\rmp,k,j} Y_{k,j} + {G}_{\rmp,k,j} U_{k,j},
      \label{eqconstraintmiddle}
\end{align}
where
\begin{align}  
    {F}_{\rmd,k,j} &\isdef 
    \setlength{\arraycolsep}{0pt}
    \renewcommand{\arraystretch}{0}
    \matl 
     \Tilde F_{\hat n,k|2,j} & \cdots   &\Tilde F_{1,k|2,j} \\
      \vdots & \ddots   &\vdots \\
      0 & \cdots   &\Tilde F_{\hat n,k|\hat n+1,j}  \\
      &0_{(\ell - \hat n)\times \hat n }&
    \matr \in\BBR^{\ell  \times  \hat n },
    \\
    {G}_{\rmd,k,j} &\isdef
    \setlength{\arraycolsep}{0pt}
    \renewcommand{\arraystretch}{0}
    \matl 
    \Tilde G_{\hat n,k|2,j} & \cdots   &\Tilde G_{2,k|2,j}  \\
     \vdots & \ddots   &\vdots  \\
      0 & \cdots   & \Tilde G_{\hat n,k|\hat n,j} \\
      &0_{(\ell - \hat n+1)\times (\hat n-1) }&
     \matr \in\BBR^{\ell  \times  \hat n-1 },
     \\
    D_k &\isdef 
    \setlength{\arraycolsep}{2pt}
    \matl y_{k-\hat n+2}& \cdots & y_{k}& \hat y_{k+1}
    & u_{k-\hat n+2} & \cdots & u_{k} \matr
    \in \BBR^{2\hat n-1},
    \\
    F_{\rmp,k,j} &\isdef
    \setlength{\arraycolsep}{2pt}
    \renewcommand{\arraystretch}{0}
    \matl
       0      & \cdots   & 0 
       & \cdots   & 0   \\
       \vdots       & \ddots   & \vdots  
       & \ddots   & \vdots  \\
     \Tilde F_{\hat n,k|\hat n +2,j}    & \cdots 
     & 0  & \cdots  & 0   \\ 
       \vdots       & \ddots   & \vdots  
       & \ddots   & \vdots  \\
       0 &\cdots &  \Tilde F_{\hat n, k|\ell+1,j} 
       &\cdots & 0   
     \matr 
     \in \BBR^{\ell \times \ell},
\end{align}
\begin{align} 
    G_{\rmp,k,j} &\isdef
    \setlength{\arraycolsep}{4pt}
     \matl
       \Tilde G_{1,k|2,j}       & \cdots   & 0 
       & \cdots   & 0   \\
       \vdots       & \ddots   & \vdots  
       & \ddots   & \vdots  \\
     \Tilde G_{\hat n,k|\hat n +1,j}    & \cdots   
     & \Tilde G_{1,k|\hat n +1,j}  & \cdots  & 0  \\ 
       \vdots       & \ddots   & \vdots  
       & \ddots   & \vdots  \\
       0 &\cdots &  \Tilde G_{\hat n, k|\ell+1,j} 
       &\cdots & \Tilde G_{1,k|\ell+1,j}   
     \matr 
     \nn\\
     &\quad 
     \in \BBR^{\ell  \times \ell }.\label{def_B_bar} 
\end{align}
Next, we rewrite \eqref{eqconstraintmiddle} as 
\begin{align}
A_{{\rm eq},k,j} \SY_{k,j} &=  b_{\rm eq,k,j},\label{x_constraints}
\end{align}
where
\begin{gather}  
    A_{\rm eq} \isdef \matl I_{\ell}- F_{\rmp,k,j} & -G_{\rmp,k,j} \matr  , \\ b_{\rm eq} \isdef    \matl -{F}_{\rmd,k,j} &{G}_{\rmd,k,j}\matr D_k .
    \label{f_constraints_matrices}  
\end{gather}
Thus, \eqref{f_cost} is subject to
\begin{gather}
    A_{{\rm eq},k,j} \SY_{} =  b_{\rm eq,k} \label{f_constraints}.
\end{gather}

\subsubsection{Optimization}
At step $k$ and at subiteration $j$, we minimize \eqref{f_cost} subject to \eqref{f_constraints} using the MATLAB quadprog function with $\SY_{k,j}$ as the initial guess.
We assume the SDC's \eqref{ABCelldef_1} and \eqref{ABCelldef_2}
are constant with respect to quadprog optimization.
Hence, the constraints \eqref{f_constraints} is linear in $\SY_{k,j}$.
Next, we define
\begin{gather}  
    Y_{k,j,{\rm opt}} \isdef \matl
    y_{k|2,j,{\rm opt}}& \hdots & y_{k|\ell+1,j,{\rm opt}} \matr^\rmT \in \BBR^{ \ell}, \\
    U_{k,j,{\rm opt}} \isdef \matl
    u_{k|1,j,{\rm opt}}& \hdots & u_{k|\ell,j,{\rm opt}} \matr^\rmT \in \BBR^{ \ell},\label{def_X_U_d}\\
    \SY_{k,j,{\rm opt}} \isdef \matl Y_{k,j,{\rm opt}}^\rmT & U_{k,j,{\rm opt}}^\rmT \matr^\rmT \in \BBR^{ 2\ell }, \label{XUstardef}
\end{gather}
where, for all $i=1,\ldots,\ell$ and $j = 0,\ldots,\nu,$ $y_{k|i+1,j,{\rm opt}}$ and $u_{k|i,j,{\rm opt}}$ are the minimizing outputs and controls at step $k$, respectively.
Lastly, we use the Broyden method as described in \cite{alhazmiIMPCACC2025} to accelerate the convergence of the subiteration and compute $U_{k|j+1}$.

Note, when the model \eqref{eq:yhat} is linear, by setting $f_i$ and $g_i$ as in subsection \ref{LPCAC}, we set $\nu = 0$, in which case IMPC is equivalent to linear PCAC (LPCAC). Otherwise, we refer to IMPC as nonlinear (NL) PCAC (NPCAC).

\section{Basis Functions}\label{sec:basis_functions}
We approximate the nonlinearity in \eqref{eq:ydef} using basis functions.
The basis functions we consider in this paper are polynomial, Fourier, and cubic Hermite splines.

\subsection{Polynomial Basis}

The polynomial approximation of 
$f\colon[a,b]\to\BBR$ is 
\begin{gather}
    f(x) \approx c_1 + c_2 x + c_3 x^2 +\cdots + c_n x^n.
\end{gather}
The basis functions are thus
\begin{gather}
    b_{{\rm p},n}(x) \isdef \matl 1 & x & x^2 & \cdots & x^n \matr^\rmT.
\end{gather}

\subsection{Fourier Basis}

The Fourier approximation of $f\colon[-L,L]\to\BBR$ is 
\begin{align}
    f(x) \approx {a_{0}} + \sum_{i=1}^{n} \left[ a_{i}\cos\frac{i\pi x}{L}+ b_{i}\sin\frac{i\pi x}{L} \right].
\end{align}
The basis functions are thus
\begin{gather}
    b_{{\rm F},n}(x) \isdef 
    \setlength{\arraycolsep}{2pt}
    \matl 1 & \cos\tfrac{\pi x}{L} & \sin\tfrac{\pi x}{L} & \cdots & \cos\tfrac{n\pi x}{L}& \sin\tfrac{n\pi x}{L} \matr^\rmT .
\end{gather}

\subsection{Cubic Hermite Spline}
We use cubic Hermite splines for piecewise interpolation.
The cubic Hermite spline approximation is $f\colon[s_0,s_{n+1}]\to\BBR$, where $n\ge1$ is the number of equal-length segments.
The nodes connecting all segments are $s_0,\ldots, s_{n+1}$.
We define the spacing $s_{\rm d} \isdef s_{2}-s_1$.
For all $i=1,\ldots, n$, we define two polynomial basis functions $p_i\colon[s_{i-1},s_{i+1}]\to\BBR$ and $m_i\colon[s_{i-1},s_{i+1}]\to\BBR$. 

For all $i = 1,\ldots, n$, the $i$-th node basis functions are given by
\begin{gather}
    b_i(x) \isdef \matl p_{i}(x)  & s_{\rm d} m_{i}(x) \matr^\rmT,
\end{gather}
where
\begin{gather}
    p_{i}(x)\isdef 
    \left\{
    \setlength{\arraycolsep}{2pt} 
    \hbox{\footnotesize
    $\begin{array}{ll}
        3t_i(x)^2 - 2t_i(x)^3, & s_{i-1} \le x < s_{i}, \\
        1- 3t_{i+1}(x)^2 + 2t_{i+1}(x)^3, & s_{i} \le x < s_{i+1}, \\
        0, & {\rm else},
    \end{array}$
    }
    \right. \\
    m_{i}(x)\isdef
    \left\{
    \setlength{\arraycolsep}{2pt} 
    \hbox{\footnotesize
    $\begin{array}{ll}
        -t_i(x)^2 + t_i(x)^3, & s_{i-1} \le x < s_{i},\\
        t_{i+1}(x)-2t_{i+1}(x)^2 + t_{i+1}(x)^3, & s_{i} \le x < s_{i+1}, \\
        0, & {\rm else},
    \end{array}$
    }
    \right. 
\end{gather}
and $t_{i}(x) \isdef \frac{x - s_{i-1}}{s_{\rm d}}$.
Therefore,
\begin{gather}
    f(x) \approx c_0 + \sum_{i=1}^n \matl c_{i,1} & c_{i,2} \matr  b_i(x) .
\end{gather}
The basis functions are thus
\begin{gather}
    b_{{\rm c},n}(x) \isdef \matl 1 & b_1(x)^\rmT & \cdots & b_{n}(x)^\rmT \matr^\rmT\in\BBR^{2n+1} .
\end{gather}

\section{Numerical Examples with LPCAC}
For the following two examples, LPCAC is applied to NL systems.
These examples are chosen to be challenging for LPCAC, thus setting the stage for NPCAC in the next section.
For these examples, we set
$f_1 = 1$ and 
$g_1 = 1$,
which makes \eqref{eq:yhat} linear, and 
$\ell = 10$.
For all examples in this and the next section, let 
$y_0 = 0.1$,
$u_0 = 0$,
$f_1 = 1$,
$\bar\theta_0 = \matl -1 & 0.01 \, \textbf{1}_{1\times g} \matr$,
$\varepsilon=1$e-$4$,
$\sigma_u = 0.01$,
$u_a = 0.1$,  
$Q = 1$, and 
the command signal $r_k \isdef 3 \sin (0.05 k)$,
while $\lambda$, $R_0$, and $R$ are tuned using a grid search for each case.

\vspace{-4pt}
\begin{example}\label{eg1}
Consider \eqref{eq:ydef}, where
\begin{gather}
    F_1 \isdef -1.1, \quad G_1(y_{k-1}) \isdef 0.9 + 0.5 \atan (y_{k-1}).\label{eq:atan}
\end{gather}
Note that \eqref{eq:ydef} with \eqref{eq:atan} is unstable.
We set 
$\lambda = 0.1$,
$R_0 = 1$, and
$R = 5$e-$1$.

Figure \ref{fig:eg1_track} shows the command-following signals.  
Figure \ref{fig:eg1_coef} compares the estimated coefficients $\widehat G_{1,k}(y_{k-1})$ with $G_1(y_{k-1})$.
%
\hfill \mbox{\large$\diamond$}
\begin{figure}
    \centering
    \includegraphics[trim = 7mm 0mm 11mm 0mm,  clip, width=0.4\textwidth]{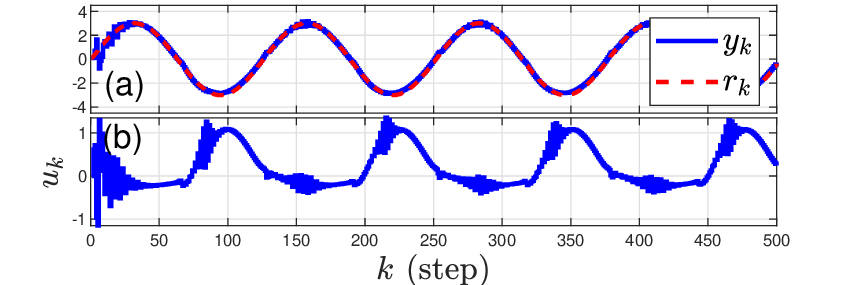}
    %
    \caption{Example \ref{eg1}: LPCAC of \eqref{eq:atan}.
    (a) shows the measurement $y_k$ and the command $r_k$;
    (b) shows the control $u_k$.
    } \label{fig:eg1_track}
\end{figure}
\begin{figure}
    \centering
    \includegraphics[trim = 1mm 0mm 2mm 0mm,  clip, width=0.4\textwidth]{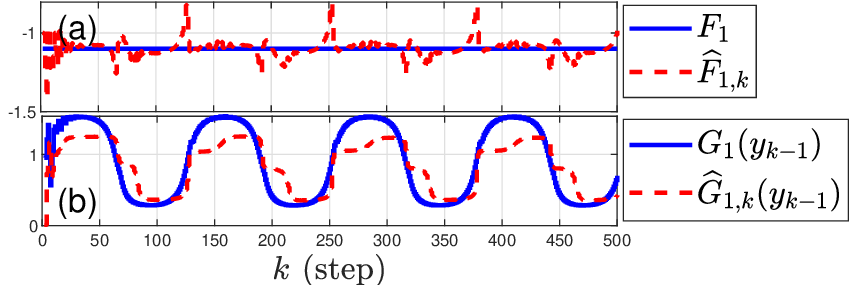}
    %
    \caption{Example \ref{eg1}: LPCAC of \eqref{eq:atan}.  
    (a) shows $F_1$ and its estimate $\widehat F_{1,k}$;
    (b) shows $G_1(y_{k-1})$ and its estimate $\widehat G_{1,k}(y_{k-1})$.
    } \label{fig:eg1_coef}
\end{figure}
\end{example}

\vspace{-4pt}
\begin{example}\label{eg3}
Consider \eqref{eq:ydef}, where
\begin{gather}
    F_1 \isdef - 1.1, \quad
    G_1(y_{k-1}) \isdef  0.4 + 0.5 \atan (y_{k-1}),\label{eq:atan_1}
\end{gather}
which is identical to \eqref{eq:atan} except that 0.9 is replaced by 0.4, in which case $G_1(y_{k-1})$ can become zero and change sign.
We set 
$R = 1$.
$\lambda$, and
$R_0$ 
are the same as Example \ref{eg1}. 

%
Figure \ref{fig:eg3_track} shows the command-following signals.  
Figure \ref{fig:eg3_coef} compares the estimated coefficients $\widehat G_{1,k}(y_{k-1})$ with $G_1(y_{k-1})$.
%
\hfill \mbox{\large$\diamond$}
\begin{figure}
    \centering
    \includegraphics[trim = 5mm 0mm 12mm 0mm,  clip, width=0.4\textwidth]{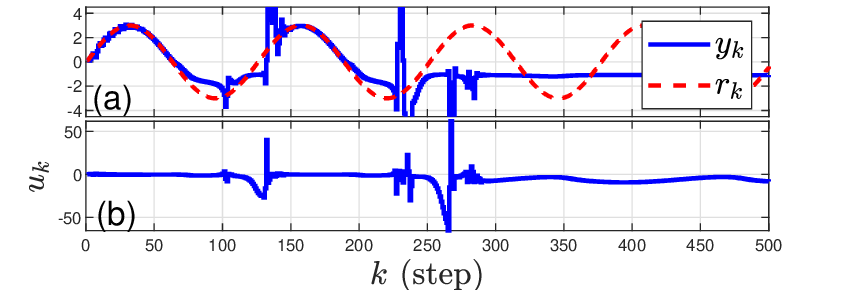}
    %
    \caption{Example \ref{eg3}: LPCAC of \eqref{eq:atan_1}.
    (a) shows the measurement $y_k$ and the command $r_k$;
    (b) shows the control $u_k$.
    Unlike Example \ref{eg1}, the controller fails to follow the command.
    } \label{fig:eg3_track}
\end{figure}
\begin{figure}
    \centering
    \includegraphics[trim =  3mm 0mm 0mm 0mm,  clip, width=0.4\textwidth]{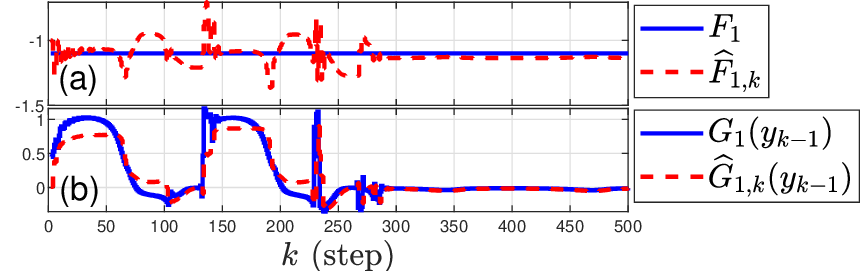}
    %
    \caption{Example \ref{eg3}: LPCAC of \eqref{eq:atan_1}. 
    (a) shows $F_1$ and its estimate $\widehat F_{1,k}$;
    (b) shows $G_1(y_{k-1})$ and its estimate $\widehat G_{1,k}(y_{k-1})$.
    Note that unlike Figure \ref{fig:eg1_coef}(b) in Example \ref{eg1}, $G_1(y_{k-1})$ crosses zero.
    } \label{fig:eg3_coef}
\end{figure}
\end{example}

It may be conjectured that the poor performance of LPCAC shown in Figure \ref{fig:eg3_track} is at least partly due to the fact that, when $G_1(y_{k-1})$ in \eqref{eq:atan_1} becomes zero, control authority is lost.
In the next section, we show that, despite the sign change in $G_1(y_{k-1})$ in \eqref{eq:atan_1}, NPCAC with section \ref{sec:basis_functions} basis in $g_1$ has much better performance than LPCAC.

\section{Numerical Examples with NPCAC}
For the following three examples, NPCAC is applied to NL systems.
As a baseline (BL) result, the NL coefficient in the NL system is included as a basis. 
The BL is compared with the $n$-element polynomial basis (PB$n$), the $n$-element Fourier basis (FB$n$), and the $n$-element cubic-spline basis (CB$n$). 
Moreover, for $k=450$, we visually compare the true NL function with the approximation.
For all subsequent examples, we set
$\nu = 10$ and 
$\ell = 20$.
%
%

\vspace{-4pt}
\begin{example}\label{eg4}
Consider \eqref{eq:atan_1}.
%
For BL, we set 
$g_{1}(y_{k-1}) =  \matl 1 &\atan y_{k-1}\matr^\rmT$,
$\lambda = 0.8$,
$R_0 = 1$e-$3$, and
$R = 0$.
For PB$2$, we set 
$g_{1}(y_{k-1}) =  b_{{\rm p},1}(y_{k-1})$,
$\lambda = 0.1$, 
$R_0 = 1$e-$3$, and 
$R = 1$e-$4$.
For FB$3$, we set 
$g_{1}(y_{k-1}) =  b_{{\rm F},1}(y_{k-1})$,
$L =6$,
$\lambda = 0.75$, 
$R_0 = 1$e-$3$, and
$R = 1$e-$3$.
For CB$3$, we set 
$g_{1}(y_{k-1}) =  b_{{\rm c},1}(y_{k-1})$,  
$s_0 = -20$,
$s_2=20$, 
$\lambda = 0.35$, 
$R_0 = 0.1$, and
$R = 1$e-$3$.

Figure \ref{fig:eg4_errors} compares the command-following error for BL, PB$2$, FB$3$, and CB$3$.  
Figure \ref{fig:eg4_yvsGGhat} compares the approximated function $\widehat{G}_{1,450}(y)$ for PB$2$, FB$3$, and CB$3$ with $G_1(y)$.  
Figure \ref{fig:eg4_coef_bar_lb} shows the estimated coefficients $\bar{F}_{1,k}$ and $\bar{G}_{1,k}$ for PB$2$.  
Figure \ref{fig:eg4_track_cs} shows the measurement $y_k$, the command $r_k$, and the control $u_k$ for CB$3$.  
Figure \ref{fig:eg4_coef_cs} compares the estimated coefficients $\widehat{F}_{1,k}$ and $\widehat{G}_{1,k}(y_{k-1})$ for CB$3$ with ${F}_{1}$ and $G_1(y_{k-1})$, respectively.
\hfill \mbox{\large$\diamond$}
\begin{figure}
    \centering
    \includegraphics[trim = 3mm 0mm 4mm 0mm,  clip, width=0.42\textwidth]{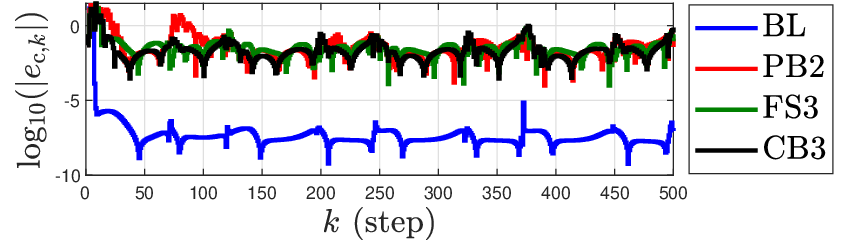}
    %
    \caption{Example \ref{eg4}: NPCAC of \eqref{eq:atan_1}.
    The plot compares the command-following error ${\rm log}_{10} (| e_{{\rm c},k}|)$  for BL, PB$2$, FB$3$, and CB$3$.
    } \label{fig:eg4_errors}
\end{figure}
\begin{figure}
    \centering
    \includegraphics[trim = 15mm 0mm 16mm 4mm,  clip, width=0.38\textwidth]{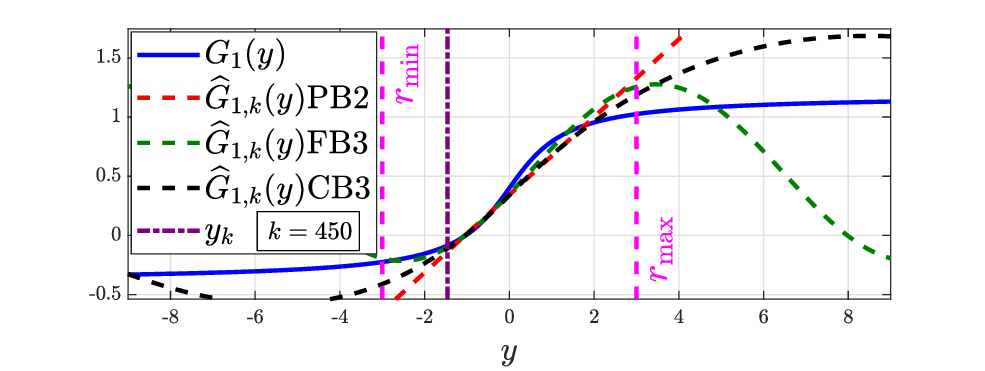}
    %
    \caption{Example \ref{eg4}: NPCAC of \eqref{eq:atan_1}.
    The plot shows $G_1(y)$ and the approximate $\widehat{G}_{1,450}(y)$ for PB$2$, FB$3$, and CB$3$.
    Note that $\widehat{G}_{1,450}(y)$ is accurate closer to $y_{450}$.
    } \label{fig:eg4_yvsGGhat}
\end{figure}
\begin{figure}
    \centering
    \includegraphics[trim = 2mm 0mm 12mm 0mm,  clip, width=0.4\textwidth]{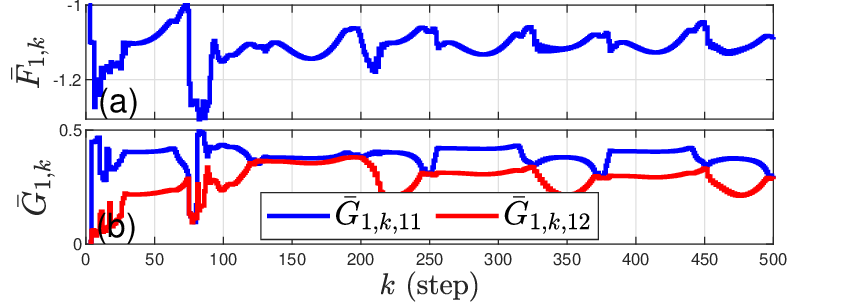}
    %
    \caption{Example \ref{eg4}: NPCAC of \eqref{eq:atan_1} for PB$2$.
    The plot shows the estimated coefficients $\bar{F}_{1,k}$ and $\bar{G}_{1,k}$ for PB$2$.
    Note that $\bar{G}_{1,k}$ for PB$2$ does not converge, indicating adaptation to $G_1(y_{k-1})$ at different step-$k$.
    } \label{fig:eg4_coef_bar_lb}
\end{figure}
\begin{figure}
    \centering
    \includegraphics[trim = 5mm 0mm 12mm 0mm,  clip, width=0.4\textwidth]{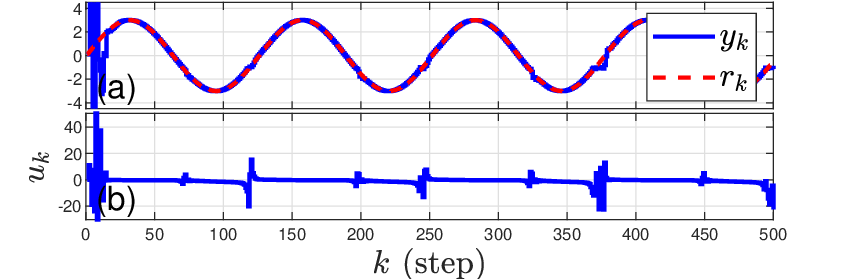}
    %
    \caption{Example \ref{eg4}: NPCAC of \eqref{eq:atan_1} for CB$3$.
    (a) shows the measurement $y_k$ and the command $r_k$;
    (b) shows the control $u_k$.
    } \label{fig:eg4_track_cs}
\end{figure}
\begin{figure}
    \centering
    \includegraphics[trim =  9mm 0mm 0mm 0mm,  clip, width=0.4\textwidth]{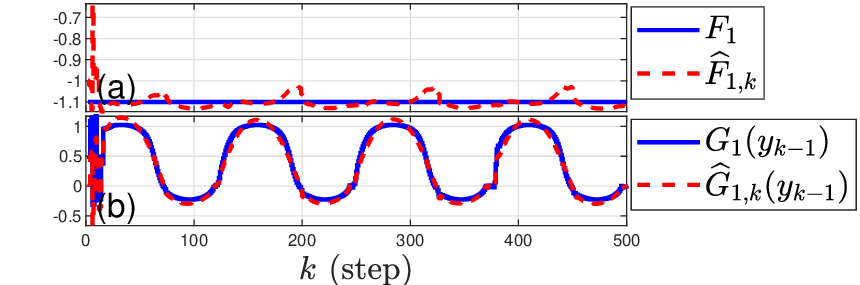}
    %
    \caption{Example \ref{eg4}: NPCAC of \eqref{eq:atan_1} for CB$3$.
    (a) shows $F_1$ and its estimate $\widehat F_{1,k}$;
    (b) shows $G_1(y_{k-1})$ and its estimate $\widehat G_{1,k}(y_{k-1})$.
    } \label{fig:eg4_coef_cs}
\end{figure}
\end{example}

\vspace{-4pt}
\begin{example}\label{eg5}
Consider \eqref{eq:ydef}, where
\begin{gather}
    F_1 \isdef - 1.1, \quad
    G_1(y_{k-1}) \isdef  0.4 + 0.5 \sin (y_{k-1}), \label{eq:sin}
\end{gather}
which is equivalent to \eqref{eq:atan_1} except that atan is replaced with sin, in which case $G_1$ is not monotonic.
%
For BL, we set 
$g_{1}(y_{k-1}) = \matl 1 &\sin y_{k-1}\matr^\rmT$.
$\lambda$,
$R_0$, and
$R$ are
the same as Example \ref{eg4}. 
For PB$2$, FB$3$, and CB$3$, 
$g_{1}$ and $L$ are
the same as Example \ref{eg4}.
For PB$2$, we set 
$R = 1$.
$\lambda$ and
$R_0$ are
the same as Example \ref{eg4}. 
For FB$3$, we set 
$\lambda=0.3$,
$R_0 = 0.01$, and
$R = 0.1$.
For CB$3$, we set 
$s_0 = -6$,
$s_2=6$,
$\lambda = 0.3$, 
$R_0 = 1$, and
$R = 0.1$.
Figure \ref{fig:eg5_errors} compares the command-following error for BL, PB$2$, FB$3$, and CB$3$.  
Figure \ref{fig:eg5_yvsGGhat} compares the approximated function $\widehat{G}_{1,450}(y)$ for PB$2$, FB$3$, and CB$3$ with $G_1(y)$.  
Figure \ref{fig:eg5_track_cs} shows the measurement $y_k$, the command $r_k$, and the control $u_k$ for CB$3$.  
Figure \ref{fig:eg5_coef_cs} compares the estimated coefficients $\widehat{F}_{1,k}$ and $\widehat{G}_{1,k}(y_{k-1})$ for CB$3$ with ${F}_{1}$ and $G_1(y_{k-1})$, respectively.
\hfill \mbox{\large$\diamond$}
\begin{figure}
    \centering
    \includegraphics[trim = 2mm 2mm 6mm 0mm,  clip, width=0.42\textwidth]{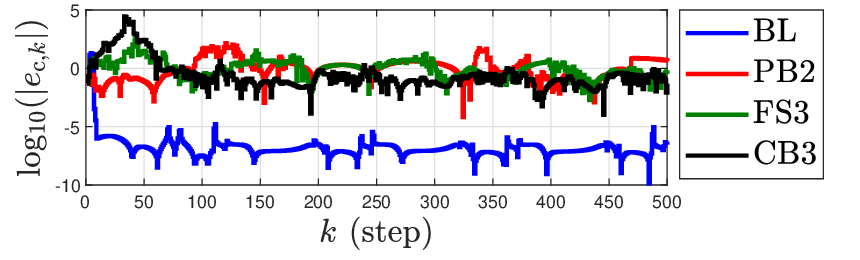}
    %
    \caption{Example \ref{eg5}: NPCAC of \eqref{eq:sin}.
    The plot compares the command-following error ${\rm log}_{10} (| e_{{\rm c},k}|)$  for BL, PB$2$, FB$3$, and CB$3$.
    } \label{fig:eg5_errors}
\end{figure}
\begin{figure}
    \centering
    \includegraphics[trim =  16mm 0mm 16mm 4mm,  clip, width=0.38\textwidth]{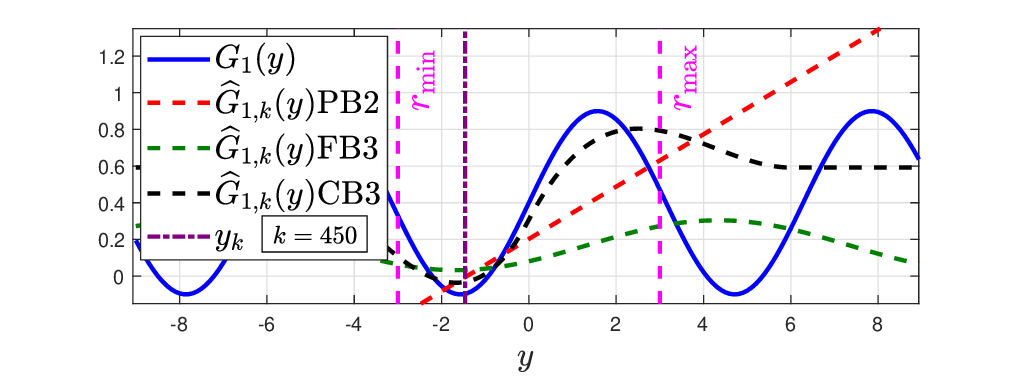}
    %
    \caption{Example \ref{eg5}: NPCAC of \eqref{eq:sin}.
    The plot shows $G_1(y)$ and the approximate $\widehat{G}_{1,450}(y)$ for PB$2$, FB$3$, and CB$3$.
    Note that $\widehat{G}_{1,450}(y)$ for CB$3$ is the most accurate around $y_{450}$.
    As a result, CB$3$ has the lowest command-following error in Figure \ref{fig:eg5_errors}.
    } \label{fig:eg5_yvsGGhat}
\end{figure}
\begin{figure}
    \centering
    \includegraphics[trim = 5mm 0mm 12mm 0mm,  clip, width=0.4\textwidth]{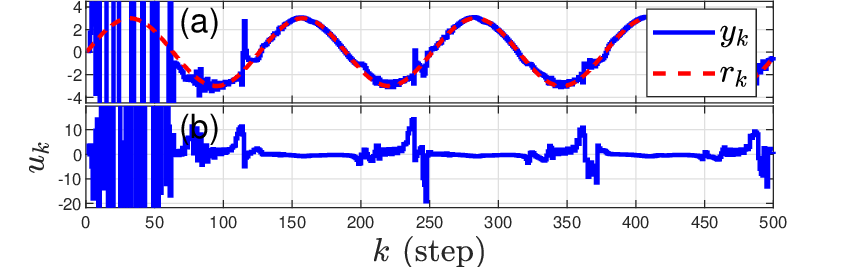}
    %
    \caption{Example \ref{eg5}: NPCAC of \eqref{eq:sin} for CB$3$.
    (a) shows the measurement $y_k$ and the command $r_k$;
    (b) shows the control $u_k$.
    %
    } \label{fig:eg5_track_cs}
\end{figure}
\begin{figure}
    \centering
    \includegraphics[trim =  10mm 0mm 0mm 0mm,  clip, width=0.4\textwidth]{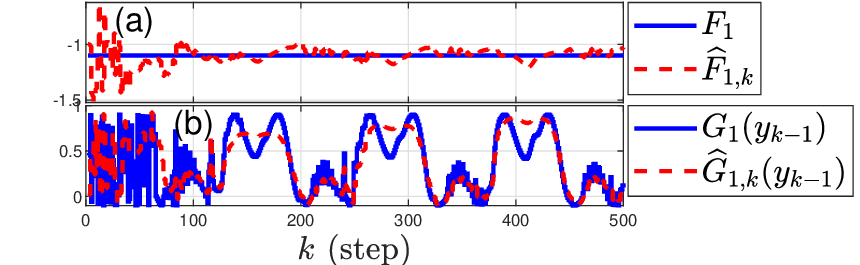}
    %
    \caption{Example \ref{eg5}: NPCAC of \eqref{eq:sin} for CB$3$.
    (a) shows $F_1$ and its estimate $\widehat F_{1,k}$;
    (b) shows $G_1(y_{k-1})$ and its estimate $\widehat G_{1,k}(y_{k-1})$.
    %
    } \label{fig:eg5_coef_cs}
\end{figure}
\end{example}

Figure \ref{fig:eg5_errors} shows that the performance for CB$3$ is better than FP$3$. 
In the next example, we increase FB$3$ to FB$5$.

\vspace{-4pt}
\begin{example}\label{eg6}
Consider \eqref{eq:sin}.
%
For BL, PB$2$, and CB$3$, 
we set
$g_{1}$, 
$R_0$, 
$\lambda$, and
$R$
the same as Example \ref{eg5}. 
For FB$5$, we set 
$g_{1}(y_{k-1}) =  b_{{\rm F},2}(y_{k-1})$
and
$R = 0.01$.
$L$, 
$\lambda$,
and
$R_0$
are the same as Example \ref{eg5}. 

Figure \ref{fig:eg6_errors} compares the command-following error for BL, PB$2$, FB$5$, and CB$3$. 
Figure \ref{fig:eg6_yvsGGhat} compares the approximated function $\widehat{G}_{1,450}(y)$ for PB$2$, FB$5$, and CB$3$ with $G_1(y)$.  
\hfill \mbox{\large$\diamond$}
\begin{figure}
    \centering
    \includegraphics[trim = 3mm 0mm 6mm 0mm,  clip, width=0.42\textwidth]{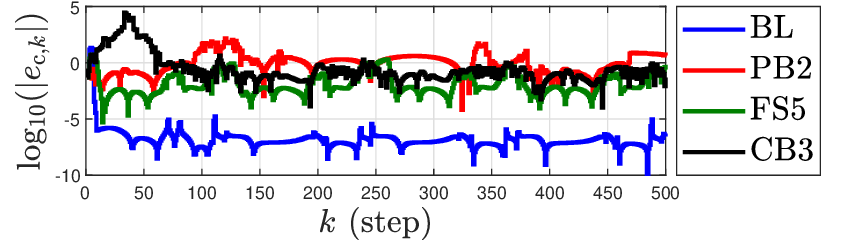}
    %
    \caption{Example \ref{eg6}: NPCAC of \eqref{eq:sin}.
    The plot compares the command-following error ${\rm log}_{10} (| e_{{\rm c},k}|)$  for BL, PB$2$, FB$5$, and CB$3$.
    } \label{fig:eg6_errors}
\end{figure}
\begin{figure}
    \centering
    \includegraphics[trim = 16mm 1mm 16mm 4mm,  clip, width=0.38\textwidth]{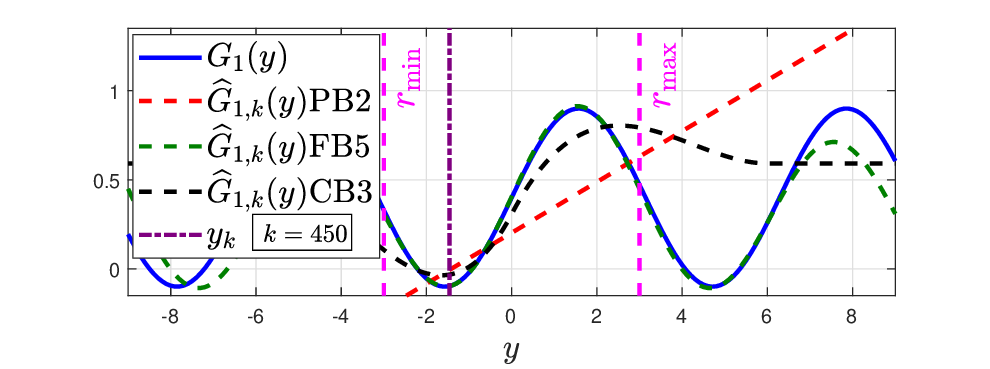}
    %
    \caption{Example \ref{eg6}: NPCAC of \eqref{eq:sin}.
    The plot shows $G_1(y)$ and the approximate $\widehat{G}_{1,450}(y)$ for PB$2$, FB$5$, and CB$3$.
    Note that $\widehat{G}_{1,450}(y)$ for FB$5$ is the most accurate around $y_{450}$ compared to Figure \ref{fig:eg5_yvsGGhat}.
    As a result, FB$5$ has the lowest command-following error in Figure \ref{fig:eg6_errors}.
    } \label{fig:eg6_yvsGGhat}
\end{figure}
\end{example}

Figure \ref{fig:eg6_errors} of Example \ref{eg6} shows an improved performance for FB$5$ compared to Figure \ref{fig:eg5_errors} of Example \ref{eg5} FB$3$. 
%
%
%

\addtolength{\textheight}{-0cm}


\section{Conclusions and Future Research}

The numerical investigation in this paper is a first step aimed at adaptive nonlinear (NL) model predictive control (MPC), where all learning is performed online without prior modeling, training, or data collection.
For a chosen set of basis functions, online system identification uses a pseudo-linear, input-output model that is linear in parameters and thus amenable to recursive least squares.

Future research will focus on several fundamental and practical issues.
First, we will examine the performance of this method on higher-order NL systems.
Next, we will investigate alternative basis functions to determine their effectiveness.
Furthermore, extension to multi-input multi-output systems is a key challenge, where the basis functions are needed for vector arguments.
Next, we will apply NL PCAC to physically motivated sampled-data systems that are not necessarily in the form of pseudo-linear input-output models.
Finally, all of these investigations can benefit from more efficient and more accurate NL MPC methods.


\bibliographystyle{IEEEtran}
\bibliography{biblio}

\end{document}